\documentclass[twocolumn]{aastex701} 

\usepackage{multirow}

\usepackage{graphics,epsf}
\usepackage[utf8]{inputenc}
\usepackage{amsmath}                
\usepackage{amsfonts}               
\usepackage{amssymb}                
\usepackage{epsfig}                 
\usepackage{graphicx}               
\usepackage{float}
\usepackage{color}
\usepackage{multirow}               

\hypersetup{
    colorlinks=true,
    linkcolor=red,   
    urlcolor=cyan}

\usepackage[colorinlistoftodos]{todonotes}

\newcommand{\km}{{~\rm km}}
\newcommand{\s}{{~\rm s}}

\newcommand{\g}{{~\rm g}}
\newcommand{\G}{{~\rm G}}
\newcommand{\K}{{~\rm K}}
\newcommand{\erg}{{~\rm erg}}
\newcommand{\yr}{{~\rm yr}}

\newcommand{\pc}{{~\rm pc}}

\newcommand{\keV}{{~\rm keV}}

\begin{document}

\title{The supernova remnant J0450.4-7050 possesses a jets-shaped point-symmetric morphology}

\author[0000-0003-0375-8987]{Noam Soker}
\affiliation{Department of Physics, Technion - Israel Institute of Technology, Haifa, 3200003, Israel; soker@technion.ac.il}
\email{soker@physics.technion.ac.il}

\begin{abstract}
By examining recently published images in different wavelengths, I identify a point-symmetric morphology in the Large Magellanic Cloud core-collapse supernova (CCSN) remnant (CCSNR) J0450.4-7050 (SNR 0450-70.9; nicknamed ``Veliki''), which I attribute to at least three pairs of energetic jets that participated in the explosion of the progenitor in the framework of the jittering jets explosion mechanism (JJEM). Two pairs of ears, a pair of blowouts in the north and south along the long axis of this SNR, and a pair of dents compose the point symmetric morphology. The fact that the symmetry axes of two pairs include pairs of opposite structural features in the inner ejecta implies that the shaping is by jets and not due to an interaction with an ambient material. While the JJEM predicts such morphologies, the competing neutrino-driven mechanism cannot account for point-symmetric morphologies. This study provides strong support for the claim that the JJEM is the primary CCSN explosion mechanism.   
\end{abstract}

\section{Introduction} 
\label{sec:intro}

The two competing theoretical explosion mechanisms that aim to explain the explosions of most core-collapse supernovae (CCSNe) are the jittering-jet explosion mechanism (JJEM; \citealt{Soker2024UnivReview, Soker2025Learning} for recent reviews) and the neutrino-driven mechanism (the delayed neutrino mechanism; \citealt{Janka2025, Janka2025Padova} for recent reviews). \cite{Soker2026G11} compares the two explosion mechanisms regarding some extra energy sources, such as a magnetar. The magnetorotational explosion mechanism (e.g., \citealt{Shibataetal2025}) is a rare case that requires a rapidly rotating pre-collapse core, and, therefore, accounts for a small fraction of CCSNe. In the framework of the JJEM, the magnetorotational explosion mechanism is an extreme case in the parameter space of the stochastic angular momentum fluctuation in the pre-collapse convective zones and the pre-collapse core rotation, corresponding to the case when the rotation is much larger than the convective velocity \citep{Soker2026G11}; still, some jittering might occur around the main angular momentum axis.      

Many observables are similar in the two explosion mechanisms, such as neutrino emission and explosion energy of most CCSNe (e.g., \citealt{Soker2025Learning}). Some observables cannot be used yet to distinguish the two mechanisms, such as gravitational waves (e.g., \citealt{Shishkinetal2025S147} for the JJEM and \citealt{Richardsonetal2025} and \citealt{Lellaetal2026} for the neutrino-driven mechanism). 
The prediction of the neutrino-driven mechanism for `failed supernovae', namely, the collapse of a massive star to a black hole without an energetic explosion and its fading, has not been confirmed yet. The recent claim (\citealt{Deetal2024, DeKetal2026}) that the fading star M31-2014-DS1 was a `failed supernova' was questioned by some studies (\citealt{Soker2024UnivReview, Beasoretal2025, Beasoretal2026}) and criticized by \cite{Soker2026Failed}. Some observations are impossible to compare between the two explosion mechanisms in a reliable manner because different neutrino-driven groups obtain embarrassingly different results, e.g., which stars explode or not (e.g., see \citealt{BoccioliFragione2024} who compare different results), and the final NS masses (e.g., \citealt{MandelMuller2020} versus \citealt{Schneideretal2023} versus \citealt{Burrowsetal2024} versus \citealt{Maltsevetal2025}).  
   
The observable property that best distinguishes between the two theoretical explosion mechanisms is the point-symmetric type morphology of CCSN remnants (CCSNRs). In the JJEM, the pairs of jets that explode the star, which the newly born neutron star (NS) launches, form point-symmetric morphologies (e.g., \citealt{Braudoetal2025}). Therefore, the JJEM predicts that many, but not all, CCSNRs possess morphologies with two or more pairs of opposite structural features that do not share the same axis. For that reason, since 2024, the research of the JJEM has focused on identifying and exploring point-symmetric morphologies in CCSNRs (e.g., some papers since 2025, \citealt{Bearetal2025Puppis, BearSoker2025, Shishkinetal2025S147, Soker2025G0901, Soker2025N132D, Soker2025RCW89, Soker2025Dust, SokerShishkin2025Vela, Soker2026G11}). Other recent studies of the JJEM conduct three-dimensional simulations of shaping by jets (e.g., \citealt{Braudoetal2025, SokerAkashi2025}), and study the convection motion in the pre-collapse core \citep{WangShishkinSoker2025} that are the seed perturbations that lead to the formation of intermittent accretion disks around the NS; the disks that launch the pairs of opposite jets that explode the star. Neutrino heating boosts the jet-driven explosion (\citealt{Soker2022nu}); however, it is not the primary energy source of the explosion.

Because the neutrino-driven explosion mechanism cannot explain all aspects of the point-symmetric morphologies of CCSNRs \citep{SokerShishkin2025Vela}, papers on this mechanism ignore these morphologies.\footnote{Studies of the neutrino-driven mechanism tend to ignore the JJEM despite its success in explaining large explosion energies and point-symmetric morphologies; the few that do mention the JJEM in recent years tend to do this in a footnote (e.g., \citealt{Janka2025, Maltsevetal2025, Antonietal2025}).}
Studies of the neutrino-driven explosion mechanism mainly simulate the revival of the stalled shock at a a radius of $\simeq 150 \km$ with neutrino heating, find the conditions for explosions, and compare simulations with some other observations beside point-symmetric morphologies (e.g., some papers since 2025: \citealt{Antonietal2025, Bambaetal2025CasA, BoccioliRoberti2025, EggenbergerAndersenetal2025, FangQetal2025, Huangetal2025, Imashevaetal2025, Laplaceetal2025, Maltsevetal2025, Maunderetal2025, Morietal2025, Nakamuraetal2025, SykesMuller2025, Orlandoetal20251987A, ParadisoCoughlin2025, Tsunaetal2025, Vinketal2025, Willcoxetal2025, Shietal2025, Mukazhanov2025, Raffeltetal2025, Vartanyanetal2025, Calvertetal2025, Giudicietal2026, LuoZhaKajino2026, Orlando2026, Rusakovetal2026, VarmaMuller2026}). 
Some studies attribute extra energy in some CCSNe to a magnetar (e.g., \citealt{Blanchardetal2026}). However, many models of magnetar energies find explosion energies of $E_{\rm exp} \gtrsim 3 \times 10^{51} \erg$ (e.g., \citealt{Aguilaretal2025, Orellanaetal2025}) that imply explosion by jets because the neutrino-driven mechanism is short of explaining such large explosion energies (e.g., \citealt{SokerGilkis2017, Kumar2025}). 
In rare cases, the NS collapses to a black hole that launches jets; this is a collapsar (e.g., \citealt{BoppGottlieb2025, Gottliebetal2025}, for recent papers). In that case, an instability due to jet-disk interaction might cause wobbling, i.e., changes in the direction of the jets (e.g., \citealt{Gottliebetal2022, Gottlieb2025}). 

The advantage of CCSNRs on CCSNe is that resolved remnants can teach us many details about the interaction with the ambient medium and emission processes (e.g., \citealt{Vuceticetal2023SerAJ, Alsaberietal2024SerAJ, Luoetal2024RAA, Filipovicetal2025PASA, Ghavametal2025SerAJ, LuXiWeietal2025RAA, Mwanikietal2025NewA, Urosevicetal2025SerAJ}), and about the explosion process, like the point-symmetric morphologies mentioned above. The search for CCSNRs with point-symmetric morphologies yielded over 15 objects. In this study, I report the identification of a jet-shaped point-symmetric morphology in the Large Magellanic Cloud SNR J0450.4-7050 (SNR 0450-709; MC 11), which \cite{Smeatonetal2025SerAJ} studied in a recent paper. They present high-resolution images of SNR J0450.4-7050 (nicknamed Veliki) that motivated this study. In Section \ref{sec:Method} I describe this CCSNR in more detail and explain the method I use to identify the point-symmetric morphology. In Section \ref{sec:Axes} I identify the main symmetry axes of this CCSNR, which compose its point-symmetric morphology. I summarize this study in Section \ref{sec:Summary} by further strengthening the JJEM as the primary explosion mechanism of CCSNe.   
\section{The method}
\label{sec:Method}

This study reveals a point-symmetric morphology in the SNR J0450.4-7050, using primarily the recently high-resolution images from \cite{Smeatonetal2025SerAJ}, who measure its size to be $150 \pc \times 81 \pc$; they analyzed some morphological aspects but not the point-symmetric morphology I identify in this study.  
Earlier studies considered this object as an SNR and presented its images.   
\cite{Mathewsonetal1985} presented H$\alpha$ and radio images, \cite{Williamsetal2004} presented in addition X-ray images (also \citealt{Maggietal2016}). \cite{Williamsetal2004} estimated a minimum age of $\approx 45,000 \yr$ and \cite{Smeatonetal2025SerAJ} estimated it as $\simeq 43,000 \yr$. \cite{Blairetal2006} reported the detection of UV emission. \cite{Cajkoetal2009} present radio images. \cite{Bozzettoetal2017} suggested that SNR J0450.4-7050 might be a CCSN, which I accept in this study.

The identification of opposite structural features follows the widely used practice of characterizing them in planetary nebula morphologies through careful visual inspection and qualitative classification (e.g., \citealt{Balick1987, Parkeretal2006, Sahaietal2007, Kwok2024}; for a preliminary, more quantitative approach, see \citealt{ShishkinMichaelis2026}). Despite being qualitative, this method has successfully led to the identification of jet-shaped morphological features  (e.g., \citealt{SahaiTrauger1998}) and their comparison with numerical simulations. This method revealed shaping processes through qualitative comparisons with simulations (e.g., \citealt{GarciaSEguraetal2022, GarciaSEguraetal2025}) and by comparing some hydrodynamical simulations with both planetary nebulae and CCSNRs (e.g., \citealt{Akashietal2018}). 
The visual inspection method for revealing jet-shaped morphologies has been applied to other systems, such as B[e] supergiants (e.g., \citealt{Kashi2023}) and luminous blue variable nebulae (e.g., \citealt{Kashi2024}).  
Most importantly, the visual method for identifying point-symmetric structures has been successfully applied to CCSNRs (e.g., \citealt{Soker2024W44, Soker2024NA1987A, Soker2026G11}).  

Several processes degrade and even destroy the point-symmetric structure that several pairs of jittering jets shape during the explosion process, including the following (e.g., \citealt{SokerShishkin2025Vela}): instabilities during the explosion process that introduce stochastic clumps and filaments; radioactive products that form hot bubbles (nickel bubbles); the natal kick of the neutron star (NS) that implies more mass ejection in the opposite direction; energy deposition by the NS remnant; interaction with a circumstellar material (CSM); interaction with the interstellar medium (ISM). In addition, in many cases, the two jets in a pair are unequal (e.g., \citealt{Soker2024CounterJet, ShishkinBearSoker2025S147}). The CSM might itself possess an axisymmetric structure, like the three rings of SN 1987A, which, at later times, introduces another symmetry axis that does not result from the explosion. When the two opposing morphological features include a high abundance of metals, the structure is clearly the result of an explosion, as these metals originate in the core. Furthermore, when the symmetry axis includes structural features in the inner ejecta, interactions with the CSM or ISM cannot explain them; they must result from the explosion process. 

In addition to these processes that degrade the point-symmetric mass distribution, others make emission intensities unequal on opposite sides of the SNR. For example, a denser CSM on one side can lead to stronger radio and X-ray emission on that side, while harder ionization radiation on one side can make some line emission stronger (e.g., of [O\textsc{iii}]).   
For all of these processes, the point-symmetric morphologies of CCSNRs are never perfect, and in some cases, only the combination of two or more emission bands reveals the two opposite sides. This is the case with the two opposite blowouts of SNR J0450.4-7050 that I discuss in Section \ref{sec:Axes}.

\section{The symmetry axes of SNR J0450.4-7050}
\label{sec:Axes}

\begin{figure}[]
	\begin{center}
\includegraphics[trim=0.0cm 11.9cm 0.0cm 0.0cm ,clip, scale=0.53]{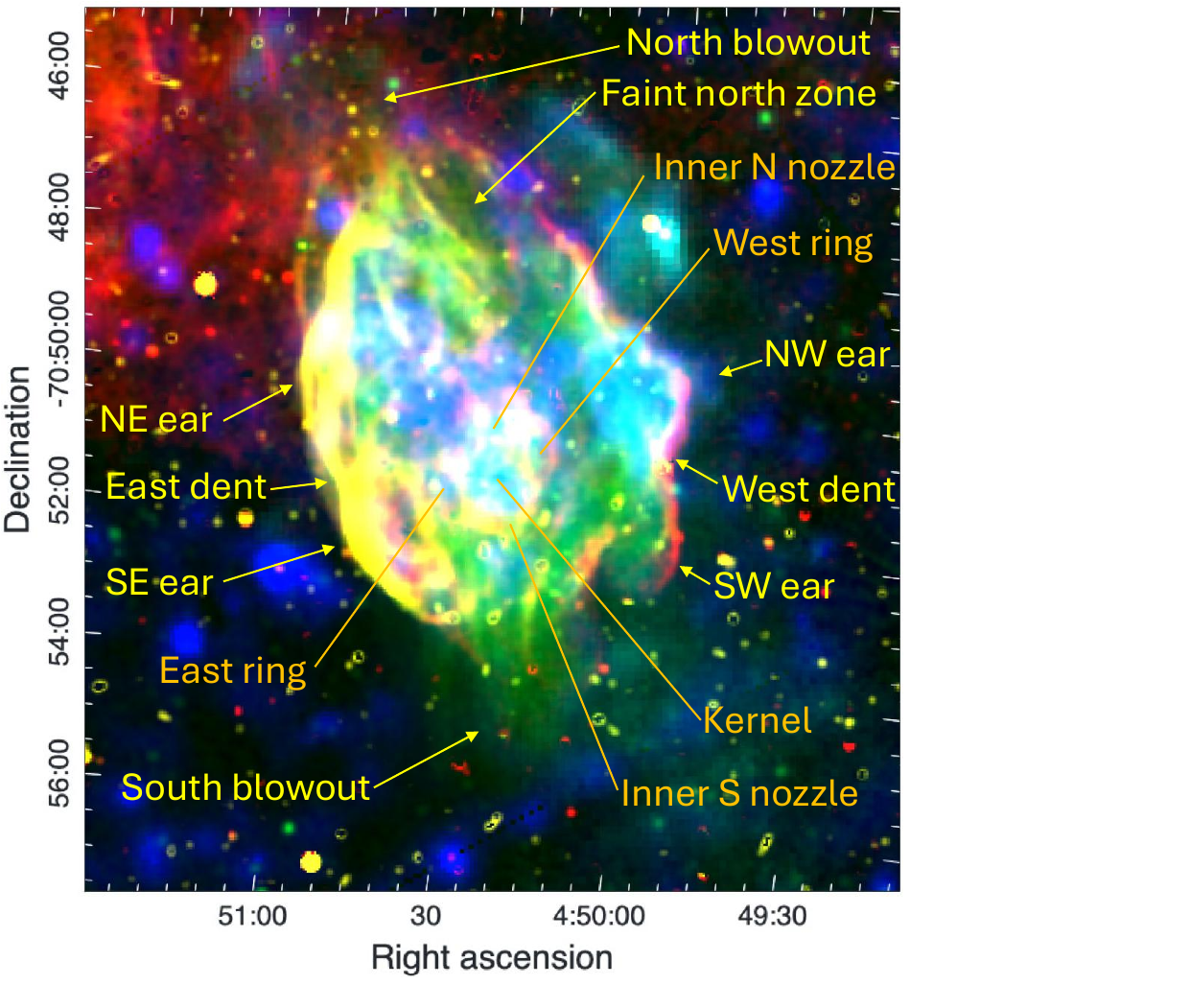} 
\caption{A five-color image of SNR J0450.4-7050 adapted from \cite{Smeatonetal2025SerAJ}. 
Red: H$\alpha$; Green: 1.3 GHz MeerKAT radio; blue: XMM-Newton $0.3-1\keV$ X-ray; yellow: [O\textsc{iii}]; cyan: $250 \mu m$ Herschel IR. 
I added the labeling of the morphological features. The kernel is the entire bright structure inside the main SNR shell.    }
\label{Fig:First}
\end{center}
\end{figure}

In Figure \ref{Fig:First}, which is a composite image of several lines and wavebands that I adapted from  \cite{Smeatonetal2025SerAJ}, I mark morphological features that I use next to identify the point symmetric structure. An `ear' is defined as a protrusion smaller than the main SNR shell that has a decreasing cross-section with distance from the center.  A nozzle is a narrow, faint zone with elongation in the radial direction that might indicate the jet's axis. A blowout is an unordered, faint protrusion from the main SNR shell. The `kernel' is the entire bright inner region, bounded by the two inner nozzles and the two rings. It is evident from the figure (and later figures) that some opposite structural features are strong in different wavebands: The north blowout is strong in H$\alpha$ and [S\textsc{ii}] while the south blowout in radio; The NW (northwest) ear is strong in H$\alpha$ and [S\textsc{ii}], while the NE (northeast) and SE (southeast) ears are strong in [O\textsc{iii}].  

\begin{figure*}[]
	\begin{center}
\includegraphics[trim=0.0cm 5.3cm 0.0cm 0.0cm ,clip, scale=0.80]{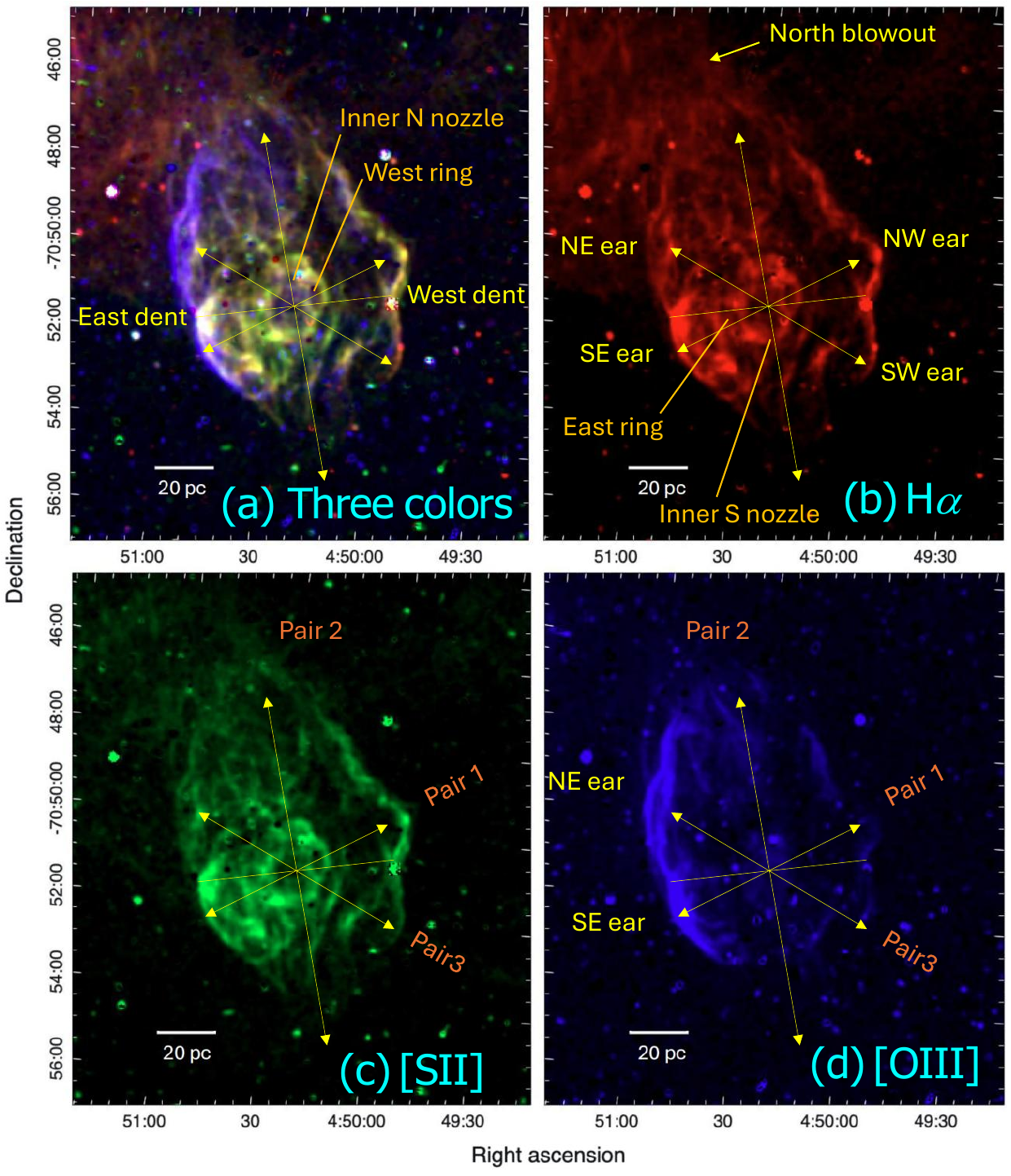} 
\caption{Images of SNR J0450.4-7050 in three lines as indicated; adapted from \cite{Smeatonetal2025SerAJ}. Panel (a) combines the other three panels. All images are linearly scaled. I added the labeling and the lines. The three double-sided arrows mark my suggestion for the axes of three pairs of energetic jets. The line connects the two dents (see Figure \ref{Fig:First}). The four lines intersect at their centers.  In all panels the point-symmetric wind-rose is of the same size.  
}
\label{Fig:4panels}
\end{center}
\end{figure*}

\begin{figure}[]
	\begin{center}
\includegraphics[trim=0.0cm 11.9cm 0.0cm 0.0cm ,clip, scale=0.52]{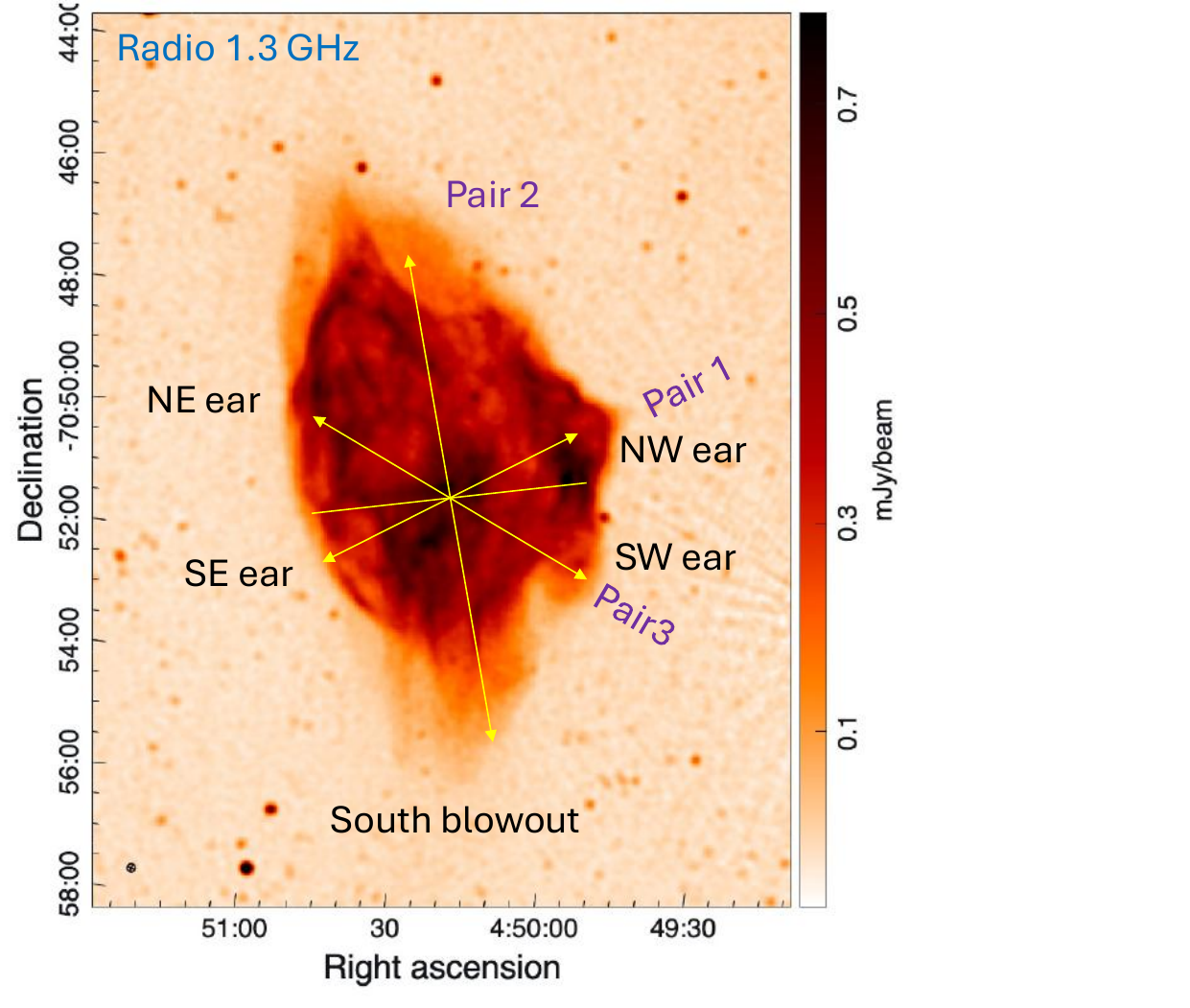} 
\caption{MeerKAT 1.3 GHz image of SNR J0450.4-7050  adapted from \cite{Smeatonetal2025SerAJ}. The image is linearly scaled. I copied the point-symmetric wind-rose from Figure \ref{Fig:4panels}. 
}
\label{Fig:Radio}
\end{center}
\end{figure}

I identify three possible jet pairs that shaped SNR J0450.4-7050; in the framework of the JJEM, more pairs are expected to have participated in the explosion process, but left no signature I can identify. I draw the jets' axes of these three pairs by two-sided yellow arrows in the four panels of Figure \ref{Fig:4panels}.  The headless line that connects the two dents is not a jet's axis, but rather a line to emphasize the point symmetrical structure. Each dent is the meeting location of two adjacent ears. The three double-sided arrows and the line intersect at their centers. In all four panels of Figure \ref{Fig:4panels}, and in Figure \ref{Fig:Radio}, the point-symmetric wind-rose is of the same relative size.
%
   
Pair 1 of jets extends from the NW ear through the west and east rings in the kernel to the SE ear. The bright X-ray zone within the NW ear (see, e.g., \citealt{Williamsetal2004, Maggietal2016} for X-ray maps) supports my claim that this ear was inflated by a jet. 

Pair 2 extends from the south blowout through the two nozzles of the kernel to the faint north zone that connects to the north blowout. Pair 2 does not point at localized structures in the far north and south, but rather at large areas. From the elongation of the SNR in that direction (north-south), I infer that Pair 2 was the most energetic pair of jets. It is possible that Pair 2 represents several pairs of jets along very similar directions, or a long-lived precessing jet. The faint north zone deserves further study. It might be the zone through which the northern jet of Pair 2 broke out to inflate the north blowout. 

The fact that Pair 1 and Pair 2 contain structures in the inner ejecta, the rings and nozzles in the kernel, respectively, implies that these axes are not due to the interaction with the CSM or the ISM. These are axes defined by two pairs of jets that participated in the explosion process. 

Pair 3 extends from the SW ear that is bright in H$\alpha$ to the NE ear that is bright in [O\textsc{iii}]. I do not identify any clear morphological feature associated with this pair in the kernel. Despite that, I consider the brightness of the NE ear in [O\textsc{iii}] emission, like the SE ear brightness, to support the claim that it was inflated by a jet. 

I can identify the three axes in the older composite image presented by \cite{Williamsetal2004}. However, only the new images by \cite{Smeatonetal2025SerAJ} allowed the clear identification of the point-symmetric morphology of SNR J0450.4-7050 and motivated this study. 

My claim that the SNR J0450.4-7050 shapes the ISM rather than the ISM shaping the remnant is compatible with the low ambient density that \cite{Smeatonetal2025SerAJ} estimates. 

\cite{Smeatonetal2025SerAJ} commented that the south blowout might be radiation leakage rather than material from the remnant, because there is no H$\alpha$ and [S\textsc{ii}] radiation from that region. The point-symmetric structure I identify here, and the H$\alpha$ and [S\textsc{ii}] filaments in the south that extend radially into the south blowout, suggest that it is a material structure.  

I consider the point-symmetric morphology of SNR J0450.4-7050, as marked by the wind-rose in Figures \ref{Fig:4panels} and \ref{Fig:Radio}, to be a robust identification. 

\section{Summary} 
\label{sec:Summary}

Analyzing high-resolution images of SNR J0450.4-7050 (nicknamed Veliki) that \cite{Smeatonetal2025SerAJ} recently published, I identified a point-symmetric morphology. It is composed of two pairs of ears, a pair of blowouts, and a pair of dents. I attribute this structure to three energetic pairs of jets; I mark the three pairs' axes by three double-sided arrows in Figures \ref{Fig:4panels} and \ref{Fig:Radio}.  The fact that Pair 1 and Pair 2 include pairs of opposite structural features in the inner ejecta, the kernel, implies that the shaping is not due to an interaction with the CSM or the ISM; these inner structures are the pair of rings and the pair of inner nozzles, respectively.  
The elongation of the SNR in the south-north direction and the large blowouts imply that Pair 2 was the most energetic pair of jets. Pair 2 might actually represent two or more pairs along very close directions that, at present, are indistinguishable. 

The robust point-symmetric morphology of SNR J0450.4-7050 is fully compatible with the JJEM and has no explanation in the competing neutrino-driven explosion mechanism. This by itself rules out the neutrino-driven mechanism as the explosion mechanism of the progenitor of SNR J0450.4-7050. But there seems to be another problem. By applying the SNR evolution model of \cite{LeahyWilliams2017}  and \cite{Leahyetal2019} to the measured X-ray temperature and emission measure from \cite{Maggietal2016}, \cite{Smeatonetal2025SerAJ} estimated the explosion energy as $E_{\rm exp} \simeq 8.6 \times 10^{51} \erg$. This explosion energy is way above what the neutrino-driven mechanism can yield ($\simeq 2 \times 10^{51} \erg$), further supporting the jet-driven explosion mechanism, i.e., the JJEM.  

As I discussed in Section \ref{sec:intro}, the observable that best distinguishes the JJEM from the neutrino-driven mechanism is the point-symmetric structure of CCSNRs. The identification of point-symmetric morphologies in close to 20 CCSNRs indicates that the JJEM is the primary explosion mechanism of CCSNRs. Future studies should reveal point-symmetric morphologies and other jet-shaped structural features in more CCSNRs. However, the number of well-resolved CCSNRs is limited. On the other hand, the number of CCSNe is rapidly growing.

There are two observables of CCSNe (before the ejecta is resolved and a jet-shaped morphology can be detected) that slowly add more support to the JJEM. The first is the high explosion energies of some CCSNe that the neutrino-driven explosion mechanism cannot account for. The second observable is the presence of two or more photospheric shells.  In the case that 2 or more strong pairs of jets are involved in the explosion process, each pair can set a strong shock wave that compresses a shell. In some of these cases, the light curve of the CCSN might reveal two or more photospheric shells. \cite{SokerShiran2025} identified two (or possibly three) photospheric shells in SN 2023ixf from the observational analysis by \cite{Zimmermanetal2024}, and \cite{ShiranSoker2026} identified two shells, at least one non-spherical, in SN 2024ggi based on observations by \cite{ChenTWetal2025}; they also discussed that these multi-shell structures are compatible with the morphologies of several jet-shaped CCSNRs.  
 
This study provides strong support for the claim that the JJEM is the primary CCSN explosion mechanism.

\section*{Acknowledgments} 
A grant from the Pazy Foundation 2026 supported this research. I thank the Charles Wolfson Academic Chair at the Technion. 

 \bibliography{reference}{}
  \bibliographystyle{aasjournal}
 
\end{document}